\documentclass[11pt]{article}
\usepackage{fullpage}                
\usepackage{graphicx}
\usepackage{amssymb}
\usepackage{epstopdf}
\usepackage{MnSymbol}
\usepackage{amsthm}
\usepackage{hyperref}
\def\e{{\rm e}}
\def\ii{{\rm i}}
\def\x{x_{\rm c}}
\def\y{y_{\rm c}}
\def\ys{y^*}
\def\be{\begin{equation}}
\def\ee{\end{equation}}

\newtheorem{cor}{Corollary}

\title{Two-dimensional self-avoiding walks and polymer adsorption: Critical fugacity estimates}

\author{Nicholas R Beaton, Anthony J Guttmann and Iwan Jensen  \vspace{3mm} \\
{\small ARC Centre of Excellence for Mathematics and Statistics of Complex Systems} \\ 
{\small Department of Mathematics and Statistics} \\ 
{\small The University of Melbourne, Victoria 3010, Australia}}

\begin{document}
\maketitle
\abstract{
Recently Beaton, de Gier and Guttmann proved a conjecture of Batchelor and Yung
that the critical fugacity of self-avoiding walks interacting with (alternate) sites on the surface of the honeycomb lattice is $1+\sqrt{2}.$ 
A key identity used in that proof depends on the existence of a parafermionic observable for 
self-avoiding walks interacting with a surface on the honeycomb lattice.  Despite the absence of a corresponding observable 
for SAW on the square and triangular lattices,  we show that in the limit of large lattices, some of the consequences observed 
for the honeycomb lattice persist irrespective of lattice. This permits the accurate estimation of the critical fugacity for the 
corresponding problem for the square and triangular lattices. We consider both edge and site weighting, 
and results of unprecedented precision are achieved. We also {\em prove} the corresponding result for the edge-weighted case for the honeycomb lattice.
 \let\thefootnote\relax\footnotetext{Email: \texttt{nbeaton, t.guttmann, i.jensen@ms.unimelb.edu.au}}
}

\section{Introduction}
Self-avoiding walks (SAW) in a half-space, originating at a site on the surface, are well-known and useful models of polymer adsorption, 
see \cite{JvR00, V98} for reviews. It is known \cite{W75} that the connective constant for such walks is the same as 
for the bulk case. To model surface adsorption, one adds a fugacity $y=\e^{\alpha}$ to sites or edges in the surface. 
Let $c_n^+(m)$ be the number of half-space walks of $n$-steps, with $m$ monomers in the surface, and define the partition function as 
$$Z_n(\alpha) = \sum_{m=0}^n c_n^+(m)\e^{m\alpha}$$
 with $\alpha = -\epsilon/k_BT,$ where $\epsilon$ is the energy associated with a surface site (or edge), $T$ is the absolute temperature 
 and $k_B$ is Boltzmann's constant. If $\epsilon$ is sufficiently large and negative, the polymer adsorbs onto the surface, while if $\epsilon$ is positive, 
 the walk is repelled by the surface. It has been shown by  Hammersley, Torrie and Whittington \cite{HTW82} in the case of the 
 $d$-dimensional hypercubic lattice that the limit $$ \lim_{n \to \infty} n^{-1}\log Z_n(\alpha)  \equiv \kappa(\alpha)$$ exists, where 
 $\kappa(\alpha)$ is the reduced, intensive, free-energy of the system. It is a convex, non-decreasing function of $\alpha,$ and therefore 
 continuous and almost everywhere differentiable. Their discussion and proof apply, {\em mutatis mutandis} to the honeycomb and triangular lattices.
 
\begin{figure}
\centering
\includegraphics[height=6cm]{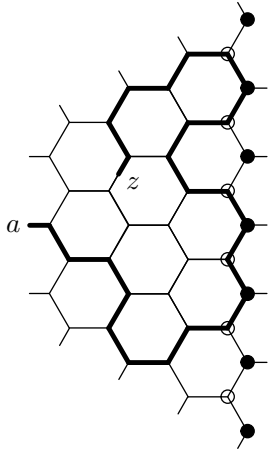}
\caption{\label{fig:exampleF}  
The figure shows the two types of surface sites on the honeycomb lattice as indicated by
solid and empty circles. A SAW starting at $a$ and finishing at $z$ is also shown. }  
\end{figure}

In the case of the honeycomb lattice there are two types of surface sites, marked as solid and empty circles
in Figure~\ref{fig:exampleF}.  In most studies just one of the the two types is weighted; 
namely, those marked with solid circles in Figure~\ref{fig:exampleF}.  In this study we also allow for a surface weight 
on the second type of sites and we study the case where all surface sites carry the same weight.

For $\alpha < 0,$ $ \kappa(\alpha) = \log \mu$ \cite{W75}, where $\mu$ is the connective constant for SAW on the given lattice. For $\alpha \ge 0,$ 
$$ \kappa(\alpha) \ge \max[\log \mu, \alpha].$$ 
This behaviour implies the existence of a critical value $\alpha_{\rm c},$ such that, for the hyper-cubic lattice, $0 \le \alpha_{\rm c} \le \log \mu.$
The situation as $\alpha \to \infty$ has only recently been rigorously established by Rychlewski and Whittington \cite{RW11}, 
who proved that $ \kappa(\alpha) $ is asymptotic to $\alpha$ in this regime. As illustrated in Figure~\ref{fig:exampleF}, 
it is convenient to attach weights $y$ to only half of the sites along the surface to allow for simplifications later on. In this case the bounds on 
$\alpha_{\rm c}$ become $0\leq \alpha_{\rm c} \leq 2\log\mu$, or equivalently $1\leq y_{\rm c} = \e^{\alpha_{\rm c}} \leq \mu^2$.

Various other quantities exhibit singular behaviour at $y_{\rm c}$. For example, the mean density of sites in the surface is given by 
$$ \rho_n(y) = \frac{1}{n}\frac{\sum_m m c_n^+(m)y^m}{\sum_m  c_n^+(m)y^m} = \frac{1}{n}\frac{\partial \log Z_n(\alpha)}{\partial \alpha}.$$ 
In the limit of infinitely long walks one has $$\rho(\alpha)=\frac{\partial \kappa(\alpha)}{\partial \alpha}.$$ From the behaviour of $\kappa$ 
given above, it can be seen that $\rho(\alpha) = 0$ for $y < y_{\rm c}$ and $\rho(\alpha) > 0$ for $y > y_{\rm c}.$ 

\section{\label{ident_honeycomb}An identity for the honeycomb lattice with a boundary}

In a recent paper Beaton, de Gier and Guttmann \cite{BdGG11} generalised a finite lattice identity by Duminil-Copin and Smirnov \cite{DC-S10} 
for the honeycomb lattice to the case where weights are introduced on alternating sites along a  boundary  as represented by
the solid circles in Figure~\ref{fig:exampleF}.  This resulted in a proof of a  conjecture of Batchelor and Yung \cite{BY95}
that the critical surface fugacity of self-avoiding walks interacting with (alternate) sites on the surface of the honeycomb lattice is $1+\sqrt{2}.$ 
Here we briefly outline the results.

Let $H$ be the set of mid-edges on a half-plane of the honeycomb lattice. We define a \emph{domain} $\Omega\subset H$ to be a 
simply connected collection of mid-edges. The set of sites adjacent to the mid-edges of $\Omega$ is denoted $V(\Omega)$. 
Those mid-edges of $\Omega$ which are adjacent to only one site in $V(\Omega)$ form $\partial\Omega$. Since surface interactions 
are the focus of this article, we will insist that at least one site of $V(\Omega)$ lies on the boundary of the half-plane.

Let $\gamma$ be a self-avoiding walk in a domain $\Omega.$
We denote by
$\ell(\gamma)$ the number of sites occupied by $\gamma$, and by $\nu(\gamma)$ the
number of contacts with the boundary.
 Define the following observable: for $a\in\partial\Omega, z\in\Omega$, set
\[
F(a,z;x,y,\sigma):=F(z) = \sum_{\gamma(a\rightarrow z) \subset \Omega} \e^{-\ii \sigma W(\gamma(a\rightarrow z))} x^{\ell(\gamma)} y^{\nu(\gamma)},
\]
where the sum is over all configurations $\gamma\subset\Omega$ for which the SAW  goes from the mid-edge $a$ to a mid-edge $z$. 
$W(\gamma(a\rightarrow z))$ is the winding angle of that self-avoiding walk. See Figure~\ref{fig:exampleF} for an example -- the SAW shown there starts on the central mid-edge of the left boundary (shown as $a$) and ends at a mid-edge $z$. As the SAW runs from mid-edge to mid-edge, it acquires a weight $x$ for each step and a weight $y$ for each contact (shown as a solid circle) with the right hand side boundary.

\begin{figure}[h]
\begin{center}
\includegraphics[height=6cm]{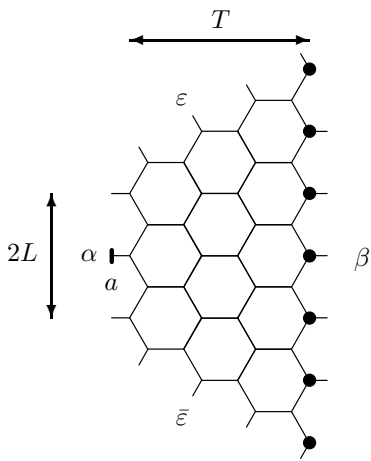}
\end{center}
\caption{\label{fig:S_boundary}
Finite patch $S_{3,1}$ of the honeycomb lattice with a boundary. }
\end{figure}

We define the following generating functions:
\begin{align*}
A_{T,L}(x,y) &:= \sum_{\genfrac{}{}{0pt}{}{\gamma\subset S_{T,L}}{a\rightarrow \alpha\slash\{a\}}} x^{\ell(\gamma)}y^{\nu(\gamma)},\\
B_{T,L}(x,y) &:= \sum_{\genfrac{}{}{0pt}{}{\gamma\subset S_{T,L}}{a\rightarrow \beta}} x^{\ell(\gamma)}y^{\nu(\gamma)},\\
E_{T,L}(x,y) &:= \sum_{\genfrac{}{}{0pt}{}{\gamma\subset S_{T,L}}{a\rightarrow \varepsilon\cup\bar{\varepsilon}}} x^{\ell(\gamma)}y^{\nu(\gamma)},
\end{align*}
where the sums are over all configurations that have a contour from $a$ to the $\alpha$, $\beta$ or $\varepsilon,\bar{\varepsilon}$  
boundaries respectively. For the SAW model in the dilute regime, the result proved in \cite{BdGG11} for the $n$-vector model simplifies (in the case $n=0$) to:

\be \label{eqn:loop_invariant}
1=\cos \left(\frac{3\pi}{8}\right) A_{T,L}(x_{\rm c},y) + \cos \left(\frac{\pi}{4}\right) E_{T,L}(x_{\rm c},y) + \frac{\ys-y}{y(\ys-1)} B_{T,L}(x_{\rm c},y).
\ee
where
\[
\ys = \frac{1}{1-2\x^2} = 1+\frac{1}{\cos(\pi/4)},\qquad\ys\x^{2}=(2)^{-1/2}.
\]

A simple corollary of \eqref{eqn:loop_invariant} is that at $y=\ys$ we have
\begin{cor}\label{cor:loop_invariant_yc}
\be \label{eqn:loop_invariant_yc}
1= \cos \left(\frac{3\pi}{8}\right) A_{T,L}(x_{\rm c},\ys) + \cos \left(\frac{\pi}{4}\right) E_{T,L}(x_{\rm c},\ys).
\ee
\end{cor}

The importance of this result is that the generating function $B_{T,L}$ for $y=\ys$ has disappeared from \eqref{eqn:loop_invariant_yc}. 
In \cite{BdGG11} we proved that the critical surface fugacity $y_c$ is equal to $\ys$. Using this result and taking the limit $L \to \infty,$ the geometry becomes 
a strip of width $T,$ and the corollary then becomes 
\be \label{eqn:strip1}
1= \cos \left(\frac{3\pi}{8}\right) A_{T}(x_{\rm c},\y) + \cos \left(\frac{\pi}{4}\right) E_{T}(x_{\rm c},\y).
\ee
In \cite{BdGG11}, we also proved that $E_T(x_c,y) = 0$ for all $0 \le y \le y_c.$ So \eqref{eqn:strip1} simplifies further to
\be \label{eqn:strip2}
1= \cos \left(\frac{3\pi}{8}\right) A_{T}(x_{\rm c},\y).
\ee
This is a remarkable equation in that it implies that $\y$ can be identified from the generating function $A_T(x_c,y)$ {\it for any width $T,$} 
simply by solving equation \eqref{eqn:strip2}. To show the power of this observation, note that virtually by inspection one can write down 
the solution for strip width $0,$ which is $$A_0(x,y) = \frac{2 x^3y}{1-x^2}. $$  Solving $1/\cos \left(\frac{3\pi}{8}\right) = A_{0}(x_{\rm c},y),$ 
recalling $x_c=1/\sqrt{2+\sqrt{2}},$ gives $y=\y = 1+\sqrt{2},$ the exact value of the critical fugacity.

\begin{figure}[h!]
\begin{center}
\includegraphics[height=10cm, angle=0]{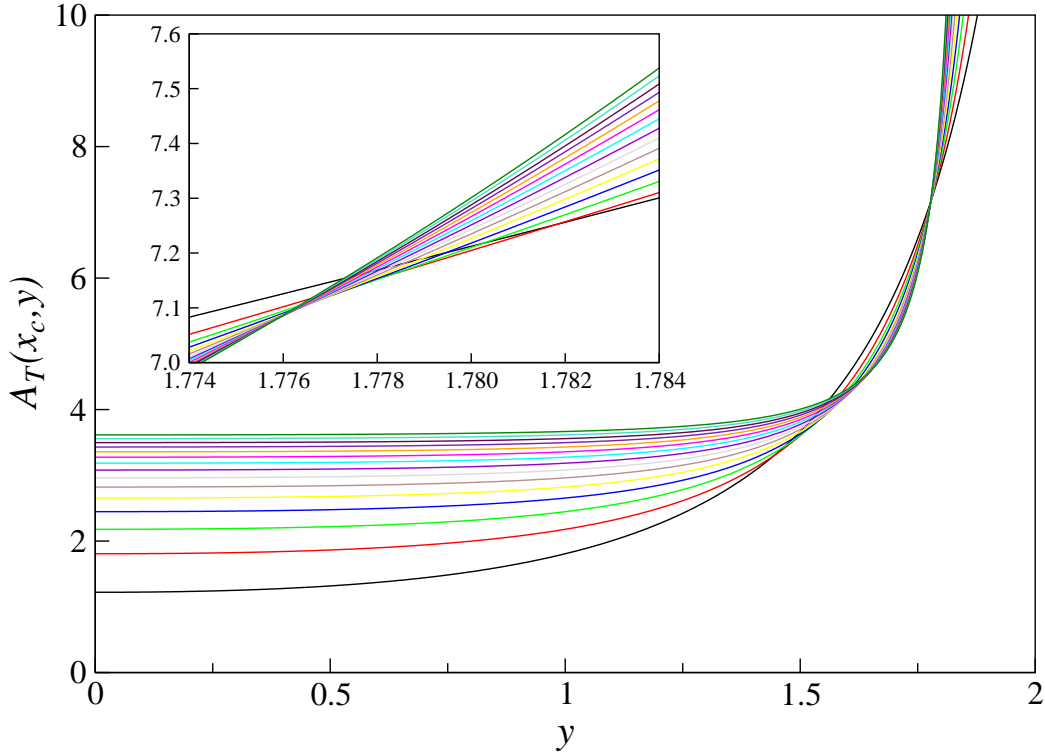}
\end{center}
\caption{\label{fig:A_sq}
Square lattice with surface site interactions. $A_T(x_c,y)$ versus $y$ for $T=1\ldots 15.$ 
Inset shows the intersection region in finer scale.}
\end{figure}

For other lattices, and even for the honeycomb lattice with interactions at every surface site, we do not have an equivalent identity, 
such as $1= c_\alpha A_{T}(x_{\rm c},\y).$ However if one plots  $A_{T}(x_{\rm c},y)$ versus $y$ in these cases, one might be 
forgiven for thinking that such an identity exists. In Figure \ref{fig:A_sq} we show a plot of $A_{T}(x_{\rm c},y)$ versus $y,$ 
for a range of strip widths $T.$ To graphical accuracy it appears that there is a unique point of intersection for plots corresponding 
to higher values of $T$. Even finer resolution, see inset, suggests that this is the case. The actual small deviation can be seen from 
the data given in Table \ref{tab:sqe}.

We denote by $y_c(T)$ the point of intersection of $A_{T}(x_{\rm c},y)$ and $A_{T+1}(x_{\rm c},y).$ We observe that the 
sequence $\{y_c(T)\}$ is a monotone function of $T.$ We argue, as in \cite{BGJ11}, that in the scaling limit all two dimensional 
SAW models are given by the same conformal field theory. Since it is known for one of these models (i.e. honeycomb lattice SAW 
with alternate site interactions) that the critical point can be found by requiring certain contour integrals to vanish, it follows that in the 
scaling limit the same should be true for all two dimensional SAW\footnote{We thank John Cardy for this observation.}. This is entirely 
consistent with our observations, and implies that $\lim_{T \to \infty} y_c(T) = y_c.$

This then suggests a potentially powerful new numerical approach to estimating $y_c.$ One calculates the generating functions 
$A_{T}(x_{\rm c},y),$ for all strip widths $T=0,1,2, \ldots T_{\rm max},$ uses these to calculate $y_c(T)$ for $T=0,1,2, \ldots T_{\rm max-1}$  
as defined above, and then extrapolates this monotone sequence by a variety of standard sequence extrapolation methods. 
A similar idea was used to furnish estimates of $x_c$ in  \cite{BGJ11}.

In Section \ref{sec:flm} we describe the derivation of the generating functions $A_T(x_c,y)$ by the finite-lattice method for a range 
of strip widths $T$ that are needed in this study. For the value of the critical step fugacity $x_c,$ we use the exact result
 $x_c = 1/\sqrt{2+\sqrt{2}}$ for the honeycomb lattice, and the best available series estimates in the case of the square and triangular lattices. 
These are $x_c({\rm sq})=0.37905227776$  \cite{JG99,IJ03},  with uncertainty in the last digit, and  $x_c({\rm tr}) = 0.2409175745$  \cite{IJ04b}, 
with similar uncertainty. We performed a sensitivity analysis of our critical surface fugacity estimates in order to determine how sensitive they 
are to uncertainties in our estimates of $x_c.$ The estimates of $x_c$ are sufficiently precise that a change in our estimate of $x_c$ by a factor of 
10 times the estimated uncertainty will not change our estimates of the surface fugacity $y_c$ in even the least significant digit.

In Section \ref{data} we estimate the critical fugacity by extrapolating  $y_c(T)$ using a range of standard extrapolation algorithms.
These are Levin's $u$-transform, Brezinskii's $\theta$ algorithm, Neville tables, Wynn's $\epsilon$ algorithm and the Barber-Hamer algorithm. 
Descriptions of these algorithms, and codes for their implementation,  can be found in \cite{G89}. However, we find the most precise estimates
are given by the Bulirsch-Stoer algorithm \cite{BS64}. This algorithm requires a parameter $w$, which can be thought of as a correction-to-scaling 
exponent. For the purpose of the current exercise, we have set this parameter to $1$, corresponding to an analytic correction, which is appropriate 
for the two-dimensional SAW problem \cite{CGJ05}. Our implementation of the algorithm is precisely as described by Monroe \cite{M02}, and 
we retained 50 digit precision throughout.

We used this method to estimate the critical fugacity for all cases of interest for two-dimensional SAWs. For the honeycomb lattice, 
discussed in Subsection \ref{honey},  we have already proved \cite{BdGG11} that $y_c = 1+ \sqrt{2}$ for the alternate site interaction model, 
as conjectured by Batchelor and Yung \cite{BY95}. It is a straightforward consequence of this result  -- the argument is given in 
Subsection \ref{honey} below -- that for the honeycomb lattice with surface edge interactions (rather than site interactions), the critical fugacity is $\sqrt{1 + \sqrt{2}}.$ 
For the honeycomb lattice site interaction problem where every surface site interacts with the walk, we find the critical fugacity to be 
$y_c = 1.46767$ where the error in this estimate (and all such estimates given below), is expected to be confined to a few parts in the 
last quoted digit. We know of no other estimate of this quantity in the literature. 

In subsection \ref{square} we discuss the critical fugacity for site and edge weighted adsorption on the square lattice. The only 
previous estimates for the site weighted case can be found in \cite{JvRR04}, where Monte Carlo methods were used to obtain 
the estimate $y_c=1.76 \pm 0.02.$ Our estimate, $y_c = 1.77564$ is three orders of magnitude more precise than this. 
For the edge weighted case, a transfer matrix estimate is given in \cite{GB89}, and is $y_c = 2.041 \pm 0.002.$ 
In \cite{GH95} a Monte Carlo estimate of comparable precision is given, $y_c = 2.038 \pm 0.002.$ 
Our estimate is $y_c = 2.040135,$ again some three orders of magnitude more precise. 

For the triangular lattice, discussed in Subsection \ref{triang} we are unaware of any previous investigations of the critical fugacity. 
We find, in Section~\ref{triang}, that $y_c({\rm site}) =  2.144181$ and $y_c({\rm edge}) =  2.950026.$ We repeat that errors in our quoted 
estimates are expected to be confined to a few parts in the last quoted digit. 

\section{Enumeration of self-avoiding walks \label{sec:flm}}

The algorithms we use to enumerate SAW interacting with a surface on 
the honeycomb, square and triangular lattices builds on the algorithm
outlined in our previous paper \cite{BGJ11} and detailed descriptions can be found
in these papers \cite{IJ04,IJ04b,IJ06}. Suffice to say that the generating functions 
for a given strip were calculated using transfer matrix (TM) techniques. 
The most efficient implementation of the TM algorithm generally involves 
bisecting the finite lattice with a boundary (this is just a line in our case) 
and moving the boundary in such a way as to build up 
the lattice site by site.  If we draw a SAW and 
then cut it by a line we observe that the partial SAW to the left of this 
line consists of a number of loops connecting two edges in the intersection, 
and at most two unconnected or free edges.  The other end of the free edge is 
an end-point of the SAW, hence there are at most two free ends. 

The sum over all contributing graphs is calculated as the boundary 
is moved through the lattice.  For each configuration of occupied or empty edges 
along the intersection we maintain a generating function $G_S$ for partial walks 
with configuration $S$. In exact enumeration studies  $G_S$ would be a 
truncated two-variable polynomial  $G_S(x,y)$ where $x$ is conjugate to the
number of steps and $y$ to the number of surface-contacts (sites or edges).  In a TM update each source 
configuration $S$ (before the boundary is moved) gives rise to a few new target configurations 
$S'$ (after the move of the boundary line) and $n=0, 1$ or 2 new edges and $m=0$ or 1 new
contacts are inserted leading to the update  $G_{S'}(x,y)=G_{S'}(x,y)+x^ny^mG_S(x,y)$.
Here we are primarily interested in the case where $A(x,y)$ or $B(x,y)$ are
evaluated at the critical point $x=x_c$. This actually makes life easier for us
since we can change to a single variable generating function $G_S(y)$ 
and update signatures  as $G_{S'}(y)=G_{S'}(y)+x_c^ny^mG_S(y)$. Here $G_S(y)$
is a  polynomial in the contact fugacity $y$   with real coefficients truncated at some
maximal degree $M$. The calculations were carried out using quadruple (or 128-bit) 
floating-point precision (achieved in FORTRAN with the REAL(KIND=16) type declaration).
 
In our calculations we truncated $A(x_c,y)$ at degree $M=1000$ and used strips
of half-length $L=M$. In Table~\ref{tab:xctest} we have listed estimates for $y_c(9)$
obtained from strips of width 10 and 9 (the crossing between  $A_{10}(x_c,y)$ and $A_9(x_c,y)$)
for various values of  $M$ and $L$. Clearly the choice $M=L=1000$ suffices to estimate
$y_c(9)$  to more than 10 digits accuracy.

 \begin{table}[htdp]
\caption{ \label{tab:xctest} The estimated value of $y_c(9)$ from the crossing
between $A_{10}(x_c,y)$ and $A_9(x_c,y)$  truncated at degree $M$ 
and using strips of half-length from $M$ up to $10M$.}
\begin{center}
\begin{tabular}{ccccc}
\hline
$M$ & $L=M$ & $L=2M$ & $L=5M$ & $L=10M$ \\ \hline
100 & 1.832547814756 & 1.778376701255 & 1.778024722094 & 1.778024722094 \\
250 & 1.776250937231 & 1.775990603337 & 1.775990594686 & 1.775990594686 \\
500 & 1.775990340341 & 1.775990291271 & 1.775990291271 & \\
1000 &1.775990291271 & & & \\
\hline
\end{tabular}
\end{center}

\end{table}%

The transfer-matrix algorithm is eminently suited for parallel
computations and here we used the approach first
described in  \cite{IJ03} and refer the interested reader to this 
publication for further detail.  The bulk of the calculations for this paper
were performed on the cluster of the NCI National Facility,
which provides a peak computing facility to researchers in Australia. 
The NCI peak facility is a Sun Constellation Cluster
with 1492 nodes in Sun X6275 blades, each containing
two quad-core 2.93GHz Intel Nehalem CPUs with most nodes
having 3GB of memory per core (24GB per node). It took a total
of about 3300 CPU hours to calculate $A_T(x_c,y)$ for $T$ up to 15 on the square lattice.
It is known \cite{IJ04} that the time and memory required to obtain the 
number of walks in a strip of width $T$  grows exponentially as $3^T$
for the honeycomb and square lattices and as $4^T$  for the triangular lattice.
So, the bulk of the time  was spent calculating $A_{15}$ and $B_{15}$,
which amounted to almost 2300 hours in the square lattice case. 
In this case we used 48 processors and the split between 
actual calculations and communications was roughly 2 to 1 (with
quite a bit a variation from processor to processor). Smaller widths
can be done more efficiently in that communication needs are lesser
and hence not as much time is used for this task.

\section{\label{data} Data analysis} 
\subsection{ \label{honey} Honeycomb lattice}
In \cite{BdGG11} we proved that the critical fugacity for the case of interactions with alternate sites on 
the honeycomb lattice is $y_c = 1 + \sqrt{2}.$ There are two other cases to consider. The first is the case 
of interactions with every surface site, and the second is the case of interactions with every edge. 
We will deal with the second case first, as it is a straightforward consequence of the proof given in \cite{BdGG11} 
that $y_c = \sqrt{1 + \sqrt{2}}$ in the second case. The proof of this result, in outline, is the following: 
We denote the generating functions $A$ and $B$, as defined in Section \ref{ident_honeycomb}, for the alternate 
site case considered in \cite{BdGG11}, by subscript ${\rm a}$ (for alternating). We denote the corresponding generating 
functions for the case with edge weighting with the subscript ${\rm e}.$ Then it is clear by inspection that 
$A_\e(x_c,y)=A_{\rm a}(x_c,y^2),$ as every time a walk contributing to the $A$ generating function passes through $n$ 
alternating surface sites, whether adjacent or not, it must pass through $2n$ surface edges. 

By the same argument,  every time a walk contributing to the $B$ generating function passes through $n$ alternating 
surface sites, whether adjacent or not, it must pass through $2n-1$ surface edges. This then gives rise to 
$B_\e(x_c,y)=\frac{1}{y}B_{\rm a}(x_c,y^2).$ From either of these two equations it follows that $y_c({\rm alternating}) = (y_c({\rm edge}))^2,$ 
hence $y_c({\rm edge})=\sqrt{1+\sqrt{2}}.$

We now consider the first case, in which every surface site carries a fugacity $y.$ We generated data for 
$A_T(x_c,y)$ for $T \le 14$ as described in Section \ref{sec:flm}, and found the intersection points where 
$A_T(x_c,y) = A_{T+1}(x_c,y), $ which defines $y_c(T).$ These data are tabulated in Table~\ref{tab:hcboth}. 
Extrapolating $y_c(T)$ as described above, we estimate $$y_c = 1.46767.$$
We also find, by an identical method of extrapolation, that $A(x_c,y_c) = 2.613,$ which is probably 
exactly $1/\cos(3\pi/8),$ as is the case when considering interactions with every alternate site, see~\eqref{eqn:strip2}.

 \begin{table}[htdp]
\caption{ \label{tab:hcboth} 
The value of $y_c(T)$ estimated from the crossing
of $A_{T}(x_c,y)$ and $A_{T+1}(x_c,y)$ for the honeycomb lattice surface site model. }
\begin{center}
\begin{tabular}{lll}
\hline
$T$ & $y_c(T)$ & $A_T(x_c,y_c(T)) = A_{T+1}(x_c,y_c(T)) $  \\ \hline
1   &1.474342684974343 &2.758023465753132 \\
2   &1.471231066324457 &2.699581979117133\\
3   &1.469859145369675 &2.671309655463187\\
4   &1.469144651946551 &2.655387366045945\\
5   &1.468728339703417 &2.645467247042683\\
6   &1.468465540675101 &2.638829094236329\\
7   &1.468289428840316 &2.634145423791235\\
8   &1.468122140755486 &2.629489693948282\\
9   &1.468008309717543 &2.626054066036805\\
10 &1.467956382495343 &2.624432487387554\\
11 &1.467915603443970 &2.623117304368586\\
12 &1.467883002922926 &2.622033892173660\\
13 &1.467856536243392 &2.621129346334020\\
\hline
\end{tabular}
\end{center}

\end{table}

\subsection{Square lattice \label{square}} 

We next consider data for the square lattice, with every surface site (vertex) carrying a fugacity $y.$ We generated data 
for $A_T(x_c,y)$ for $T \le 15$ as described in Section \ref{sec:flm}, and found the intersection points where  
$A_T(x_c,y) = A_{T+1}(x_c,y), $ which defines $y_c(T).$ These data are tabulated in Table~\ref{tab:sqs}. 
Extrapolating $y_c(T)$ as described above, we estimate $$y_c = 1.77564.$$
We also find, by an identical method of extrapolation, that $A(x_c,y_c) = 2.678405,$ which is  $1.024981/\cos(3\pi/8).$ 
In \cite{BGJ11} we found, for the non-interacting case (corresponding to $y=1$), $A(x_c,1) = 2.678365=1.024966/\cos(3\pi/8).$ 
Thus there appears to be a very weak $y$ dependence. (In the normalization of the generating function $A_T(x_c,y)$ used here, 
two extra half-steps are included, giving an extra factor of the step fugacity $x_c,$ compared to the value that would be quoted 
if contributing walks started and ended {\it on} the surface. This explains the difference between the values quoted in 
Table~\ref{tab:sqs} and the ordinates in Figure~\ref{fig:A_sq}.)  

 \begin{table}[htdp!]
\caption{ \label{tab:sqs} 
The value of $y_c(T)$ estimated from the crossing
of $A_{T}(x_c,y)$ and $A_{T+1}(x_c,y)$ for the square lattice surface site model.}
\begin{center}
\begin{tabular}{lll}
\hline
$T$ & $y_c(T)$ & $A_T(x_c,y_c(T)) = A_{T+1}(x_c,y_c(T)) $  \\ \hline
 1  &1.781782909906119 &2.748677355944862 \\
 2  &1.778386591113354 &2.715115253913871\\
 3  &1.777378005442640 &2.704018907440273\\
 4  &1.776850407093364 &2.697681121136133\\
 5  &1.776527700942633 &2.693512738663579\\
 6  &1.776316359764735 &2.690608915840792\\
 7  &1.776170974231462 &2.688500944397294\\
 8  &1.776066934443028 &2.686918847615982\\
 9  &1.775990033953699 &2.685698355993929\\
10 &1.775931645420429 &2.684735010917280\\
11 &1.775886299456907 &2.683959815456866\\
12 &1.775850398954429 &2.683325675630414\\
13 &1.775821502307431 &2.682799521958416\\
14 &1.775797906369155 &2.682357553489197\\
\hline
\end{tabular}
\end{center}

\end{table}

Table~\ref{tab:sqe} shows the corresponding data for the edge-weighted case. 
Extrapolating $y_c(T)$ as described above,  we estimate 
$$y_c = 2.040135.$$ 
We also find that $A(x_c,y_c) = 2.678405,$ which is  $1.024981/\cos(3\pi/8).$ In \cite{BGJ11} we found, 
for the non-interacting case (corresponding to $y=1$), $A(x_c,1) = 2.6783=1.0249/\cos(3\pi/8).$ 
This is too imprecise to see any evidence of $y$ dependence.
 \begin{table}[htdp!]
\caption{\label{tab:sqe}
The value of $y_c(T)$ estimated from the crossing
of $A_{T}(x_c,y)$ and $A_{T+1}(x_c,y)$ for the square lattice surface edge model.  }
\begin{center}
\begin{tabular}{lll}
\hline
$T$ & $y_c(T)$ & $A_T(x_c,y_c(T)) = A_{T+1}(x_c,y_c(T)) $  \\ \hline
1  &2.023317607727152  &2.519464246890523\\
2  &2.031649211433080  &2.585125356952430\\
3  &2.035085448834840  & 2.616332757155513\\
4  &2.036771224259312  &2.633293109539552\\
5  &2.037723730407517  &2.643677266387231\\
6  &2.038317002192238  &2.650588857893349\\
7  &2.038712823877066  &2.655469267857106\\
8  &2.038990695898482  &2.659069610531442\\
9  &2.039193569770578  &2.661816780067225\\
10 &2.039346383471084 &2.663969985883853\\
11 &2.039464457297598 &2.665695001241074\\
12 &2.039557641399558 &2.667102372510593\\
13 &2.039632511102958 &2.668268404182947\\
14 &2.039693596208206 &2.669247312794744\\
\hline
\end{tabular}
\end{center}

\end{table}

\subsection{Triangular lattice} \label{triang}

We next consider data for the triangular lattice, with every surface site (vertex) carrying a fugacity $y.$ 
We generated data for $A_T(x_c,y)$ for $T \le 11$ as described in Section \ref{sec:flm}, and found the 
intersection points where  $A_T(x_c,y) = A_{T+1}(x_c,y), $ which defines $y_c(T).$ These data are tabulated in 
Table~\ref{tab:sqs}. Extrapolating $y_c(T)$ as described above, we estimate $$y_c = 2.144181.$$
We also find, by an identical method of extrapolation, that $A(x_c,y_c) = 4.97002,$ which is  
$1.901944/\cos(3\pi/8).$ In \cite{BGJ11} we found, for the non-interacting case (corresponding to $y=1$), 
$A(x_c,1) = 4.970111=1.901979/\cos(3\pi/8).$ Thus there again appears to be a very weak $y$ dependence.
 \begin{table}[htdp!]
 \caption{\label{tab:trs}
The value of $y_c(T)$ estimated from the crossing
of $A_{T}(x_c,y)$ and $A_{T+1}(x_c,y)$ for the triangular lattice surface site model. }
\begin{center}
\begin{tabular}{lll}
\hline
$T$ & $y_c(T)$ & $A_T(x_c,y_c(T)) = A_{T+1}(x_c,y_c(T)) $  \\ \hline
1  &2.169017975620833 &5.299883162257977\\
2  &2.152124186067447 &5.089804987842667\\
3  &2.147952081330057 &5.033100087535114\\
4  &2.146325209334416 &5.009022287728647\\
5  &2.145537862947824 &4.996485228732837\\
6  &2.145102964455591 &4.989109337635192\\
7  &2.144840361941141 &4.984402909686655\\
8  &2.144671215263562 &4.981219362650799\\
9  &2.144556764080381 &4.978968525942606\\
10&2.144476246964690 &4.977320728801566\\
\hline
\end{tabular}
\end{center}

\end{table}

Table~\ref{tab:tre} shows the corresponding data for the edge weighted case. 
Extrapolating $y_c(T)$ as described above, we estimate 
$$y_c =  2.950026.$$
We also find that $A(x_c,y_c) = 4.9696,$ which is  $1.90178/\cos(3\pi/8).$ In \cite{BGJ11} we found, 
for the non-interacting case (corresponding to $y=1$), $A(x_c,1) = 4.970111=1.901979/\cos(3\pi/8).$ 
Again, there is evidence of weak $y$ dependence.
 \begin{table}[htdp!]
\caption{ \label{tab:tre}
The value of $y_c(T)$ estimated from the crossing
of $A_{T}(x_c,y)$ and $A_{T+1}(x_c,y)$ for the triangular lattice surface edge model. }
\begin{center}
\begin{tabular}{lll}
\hline
$T$ & $y_c(T)$ & $A_T(x_c,y_c(T)) = A_{T+1}(x_c,y_c(T)) $  \\ \hline
1  &2.933665548671216 &4.793416679321919\\
2  &2.939352607034002 &4.841229819027843\\
3  &2.942788011875285 &4.873934294210283\\
4  &2.944814166604381 &4.895179517868169\\
5  &2.946090146548846 &4.909648090189844\\
6  &2.946944189466541 &4.919989731979732\\
7  &2.947544335340955 &4.927679988442194\\
8  &2.947982663246637 &4.933582932189477\\
9  &2.948312910101248 &4.938231892866670\\
10&2.948568146735367 &4.941971526310544\\
\hline
\end{tabular}
\end{center}

\end{table}

\section{Conclusion}

We have estimated the critical fugacity for surface adsorption for two-dimensional SAW on all regular lattices 
for both the case of site and edge interactions. Many of these estimates are new. Those that are not are several 
orders of magnitude more precise than pre-existing estimates. Uniquely for the case of the honeycomb lattice 
with edge interactions, we give the exact value of the critical fugacity, and also prove it. 
Our results are summarised in Table~\ref{tab:sum}.

 \begin{table}[htdp!]
 \caption{ \label{tab:sum}
Estimated critical fugacity $y_c$ for surface adsorption. }
\begin{center}
\begin{tabular}{lll}
\hline
Lattice & Site weighting &Edge weighting  \\ \hline
Honeycomb & 1.46767 & $\sqrt{1 + \sqrt{2}}$ \\
Square & 1.77564& 2.040135 \\
Triangular & 2.144181& 2.950026\\
\hline
\end{tabular}
\end{center}

\end{table}

\section*{Acknowledgements}
AJG and IJ acknowledge financial support from the Australian Research Council. NRB was supported by the ARC Centre of Excellence for Mathematics and Statistics of Complex Systems (MASCOS).
This work was supported by an award under the Merit Allocation Scheme 
on the NCI National Facility at the ANU. 

\end{document}